\documentclass[12pt]{article}
\usepackage{color,graphicx,amsfonts,amsmath,amssymb,mathrsfs,comment}
\title{Reality and the Role of the Wavefunction in Quantum Theory}
\author{
Sheldon Goldstein\footnote{Departments of Mathematics, Physics and Philosophy,
     Rutgers University, Hill Center, 
     110 Frelinghuysen Road, Piscataway, NJ 08854-8019, USA.
     E-mail: oldstein@math.rutgers.edu}  \ and
Nino Zangh\`\i\footnote{Dipartimento di Fisica, Universit\`a di
    Genova and INFN sezione di Genova, Via Dodecaneso 33, 16146
    Genova, Italy.  E-mail: zanghi@ge.infn.it}
}
\date{January 19, 2011}

\newcommand{\bc}{\begin{center}}
\newcommand{\ec}{\end{center}}
\definecolor{gray}{gray}{.4} 
\definecolor{lightgray}{gray}{.7}
\renewcommand{\Psi}{\text{$\varPsi$}}
\newcommand{\OQT}{orthodox quantum theory}
\newcommand{\Sc}{Schr\"odinger}
\newcommand{\rvarn}[1]{\ensuremath{\mathbb{R}^{#1}}}
\newcommand{\dd}{\ensuremath{3}}
\newcommand{\nrd}{\ensuremath{{}^N\mspace{-1.0mu}\rvarn{\dd}}}

\newcommand{\z}[1]{{\bf #1}}
\begin{document}
\maketitle
\begin{abstract}
 The most puzzling issue in the foundations of quantum mechanics is perhaps that of the status of the wave function of a system in a quantum universe. Is the wave function objective or subjective? Does it represent the physical state of the system or merely our information about the system? And if the former, does it provide a complete description of the system or only a partial description? We shall address these questions here mainly from a Bohmian perspective, and shall argue that part of the difficulty in ascertaining the status of the wave function in quantum mechanics arises from the fact that there are two different sorts of wave functions involved. The most fundamental wave function is that of the universe. From it, together with the configuration of the universe, one can define the wave function of a subsystem. We argue that the fundamental wave function,  the wave function of the universe, has a law-like character. 
\end{abstract}

\section{Questions About the Wave Function} 
We shall be concerned here with the role and status of the wave function in quantum theory, and especially in Bohmian mechanics. What we shall describe is joint work with Detlef D\"urr.

The wave function is arguably the main innovation of quantum theory. Nonetheless the issue of its status has not received all that much attention over the years. 
A very welcome step is thus this very book, whose main concern is the question, What the hell is this strange thing, the wave function, that we have in quantum mechanics? What's going on with that? Who ordered that?  

In more detail, is the wave function subjective or epistemic, or is it objective? Does it merely describe our  information or does it describe an observer independent reality? What's the deal with collapse? Why does the wave function collapse? What's going on there? And if the wave function is objective, is it some sort of concrete material reality or something else?

Let us say a word about what it means  for the wave function to be merely epistemic. To us that means first that there is something else, let's call it $X$, describing some physical quantity, say the result of an experiment, or maybe the whole history up to the present of some variable or collection of variables---things we're primarily interested in. And then  to say that the wave function is merely epistemic is to say that it is basically equivalent to a probability distribution on the space of possible values for $X$.

You should note that \OQT\ is not of this form. That's  because the $X$ is in effect a hidden variable and there are no hidden variables in  \OQT---there's just the wave function. So the wave function is certainly not merely epistemic in  \OQT.

And neither is Bohmian mechanics \cite{Bohm52,Bell66,Gol01} of this form. In Bohmian mechanics it is indeed the case that the wave function sort of has a probabilistic role to play, since the absolute square of the wave function gives the probability of the configuration of the Bohmian system. However, that's not the only role for the wave function in Bohmian mechanics; it's  not its fundamental role  and certainly not its most important role.

We should all agree---and maybe this is the only thing we would all agree upon---that there are three possibilities for the wave function: (i) that it is everything, as would seem to be the case with Everett \cite{Eve57}, (ii) what would seem to be the most modest possibility, that it is something (but not everything), as with Bohmian mechanics, for example, where there's the wave function and something else, or (iii) maybe it's nothing---that would solve the problem of what's this weird thing,  the wave function: if you can get rid of it you don't have to agonize about it.

\section{Bohmian Mechanics} %
Let's turn to Bohmian mechanics, for us the simplest version of quantum mechanics. In Bohmian mechanics you've got  for an $N$-particle system the usual quantum mechanical wave function $\psi=\psi(\mathbf q_{_1}, \dots,  \mathbf q_{_N})$---in the simplest case, of spin-0 particles, a complex-valued function of the ``generic positions'' of the particles---but it's not everything: in addition to the wave function you've got the actual positions of the particles,  $\mathbf Q_{_1}, \dots,  \mathbf Q_{_N}$, which form the configuration $Q$.\footnote{We use lower case letters, such as $\mathbf q_{_1}, \dots,  \mathbf q_{_N}$, for generic position and configuration variables in quantum theory, reserving capitals for the actual positions and configurations. It is interesting that in \OQT\ one also has  generic position variables as arguments of the wave function even though there are no (unmeasured) actual  positions. We find this rather odd.}

We say that the positions of the particles provide the primitive ontology of the theory 
\cite{DGZ92,Gol98,AGTZ06}. In so saying we wish to convey that the whole point of the theory---and the whole point of the wave function---is to define a motion for the particles, and  it's in terms of this motion that pointers end up pointing and experiments end up having results, the kinds of  results that it was the whole point of quantum mechanics to explain. So the connection to physical reality in the theory is  via what we're calling the primitive ontology of the theory, in Bohmian mechanics the positions of the particles.

The wave function would seem to be part of the ontology. It's real  in that sense. It's not subjective in Bohmian mechanics---it has a rather real role to play: it has to govern the motion of the particles. But it's not part of the primitive ontology. Bohmian mechanics is fundamentally about particles and their motions, and not wave functions.

Here, for the record, are the equations of Bohmian mechanics. First of all, you've got for the wave function the usual Schr\"odinger equation
\begin{equation}\label{se}
 i\hbar\frac{\partial \psi}{\partial t} = H\psi \,,
\end{equation}
with the usual Schr\"odinger Hamiltonian
$$
H=-\sum_{k=1}^N\frac{\hbar^2}{2m_k}\nabla^2_k+V\,.
$$
Here $\nabla_k=(\partial/\partial x_k,\partial/\partial y_k,\partial/\partial z_k)$ is the position-gradient for the $k$-th particle, and $V=V(q)$ is a real-valued function of the configuration called the potential energy function.

The only thing Bohmian mechanics adds---in addition to the positions of the particles as actual variables in the theory to be taken seriously, and not just talked about in connection with measurements---is an equation of motion for the positions:
\begin{equation}\label{ge}
  \frac{d\mathbf Q_k}{dt}=\frac{\hbar}{m_k}
  \mbox{Im} \frac{\psi^{*}\mathbf{\nabla}_k \psi}{\psi^{*}\psi}(\mathbf
  Q_{_1}\ldots,\mathbf Q_{_N})\,.
\end{equation}
This equation, the new equation in Bohmian mechanics, is kind of an obvious equation. It is more or less the first thing you would guess if you asked yourself, What is the simplest motion of the particles that could reasonably be defined in terms of the wave function? (It however may not look so obvious.)

It might seem a bit pointless to have $\psi^{*}$ in both the numerator and the denominator of \eqref{ge}, so that it cancels,  leaving just $\mbox{Im}(\mathbf{\nabla}_k\psi/\psi)$ times a prefactor, which is the same as $\nabla_k S/m_k$, the more familiar  way of writing the right hand side---the velocity field in Bohmian mechanics. There are two good reasons for writing it in the above apparently more complicated form. 

In this way the formula  makes sense automatically for particles with spin, for which we would need wave functions with many components, instead of the simple single-component complex-scalar valued wave functions appropriate for systems of particles without spin (spin-0 particles). For example, the wave function for a system of $N$ spin-1/2 particles has $2^N$ components: the value of $\psi$ at a given configuration $Q$ is given by $2^N$ complex numbers, and not just one. In such a case it is not clear what    $\mbox{Im}(\mathbf{\nabla}_k\psi/\psi)$ could possibly mean. Why should the $x$-component of this, like the $x$-component of the velocity of a particle, be a scalar? What could even be meant by the ratio of two multi-component objects? 

But if you interpret the products involving $\psi$s in the numerator and denominator in the formula above as ``spinor inner products,'' involving sums of products of components of $\psi$---and that's the natural way to understand such expressions, the natural product to form with such multi-component wave functions---then the very same formula that is valid for particles without spin remains valid for particles with spin, providing an equation of motion for such particles that does exactly what you want it to do. It works perfectly. So you don't need to do anything extra in Bohmian mechanics to deal with spin. That's one reason for writing the equation as above.

The other reason is that the denominator, $\psi^{*}\psi$, is the familiar quantum probability density $\rho$, and the numerator the quantum probability current $J_k$, so the right hand side is $J_k/\rho$, which is first of all a fairly obvious thing to guess for a velocity, and second of all, because the velocity is $J_k/\rho$, the $|\psi|^2$--probabilities play the role they do in Bohmian mechanics.

As a consequence of this role, the usual quantum randomness emerges. One obtains  the {\em quantum equilibrium hypothesis,} that whenever a system has wave function $\psi$, its configuration is random,  with distribution given by $|\psi|^2$. Exactly what this means and how this comes is a long and controversial story, which we shall not go into here. But using the quantum equilibrium hypothesis one can establish the empirical equivalence between Bohmian mechanics and orthodox quantum theory, including the emergence of operators as ``observables'' and the collapse of the wave packet \cite{DGZ92,DGZ04}.

\section{The Wave Function of a Subsystem}\label{cwf}

A crucial ingredient in the extraction of the implications of Bohmian mechanics is the notion of the wave function of a subsystem of a Bohmian universe, a universe of  particles governed by the equations of Bohmian mechanics, defining a motion choreographed by  the {\em wave function of the universe} $\Psi$.

In almost all applications of quantum mechanics it is the wave function of a subsystem with which we are concerned, not the wave function of the universe. The latter, after all, must be rather elusive. Most physicists don't deal with the universe as a whole. They deal with subsystems more or less all the time: a hydrogen atom, particles going through Stern-Gerlach magnets, a Bose-Einstein condensate, or whatever. And yet from a fundamental point of view the only genuine Bohmian system in a Bohmian universe---the only system you can be sure is Bohmian---is the universe itself, in its entirety. It can't be an immediate consequence of that that subsystems of a Bohmian universe are themselves Bohmian, with the motion of their particles governed by wave functions in the Bohmian way. 

That is, one can't simply demand of subsystems of a Bohmian universe that they be Bohmian systems in their own right. The behavior of the parts of a big system are already determined by the behavior of the whole. And what you have for the whole is the wave function $\Psi$ of the universe,  together with its configuration $Q$.  That's your data. That's what objective in a Bohmian universe. The wave function of a subsystem, if it exists at all, must be definable in terms of that data. 

Now corresponding to a subsystem of the universe is a splitting $Q=(Q_{sys},Q_{env})=(X,Y)$ of its configuration $Q$ into the configuration $Q_{sys} = X$ of the subsystem, the ``$x$-system,'' formed from the positions of the particles of the subsystem, and the configuration $Q_{env}=Y$ of the environment of the subsystem---the configuration of everything else. So the data in terms of which the wave function of a subsystem must be defined are the universal wave function $\Psi(q)=\Psi(x,y)$ and the actual configurations $X$ of the subsystem and $Y$ of its environment. 

The first guess people make about what the wave function $\psi(x)$ of the  $x$-subsystem should be usually turns out to be wrong. The right guess, and the natural thing to do, is to define the wave function of a subsystem in this way: Remembering that the wave function of a subsystem should be a function on its configuration space, a function, that is, of $x$ alone, you take the universal wave function $\Psi(x,y)$ and plug the actual configuration $Y$ of the environment into the second slot to obtain a function of $x$, \begin{equation}\label{cwf1}
\psi(x)=\Psi(x,Y)\,.
\end{equation}

If you think about it you see that this is exactly the right definition. The situation is simplest for spin-0 particles, which we will henceforth assume. First of all, it is easy to see that the velocity that the configuration $X$ of the subsystem inherits from the motion of the configuration $Q$ can be expressed in terms of this $\psi$ in the usual Bohmian way. In other words, if $dQ/dt=v^{\Psi}(X,Y)$ then $dX/dt=v^{\psi}(X)$ for $\psi(x)=\Psi(x,Y)\,.$ 

However, the evolution law for the wave function of the subsystem need not be Bohmian:
Explicitly putting in the time dependence, we have for the wave function of the $x$-system at time $t$:
$$\psi_t(x)=\varPsi_t(x,Y_t)\,.$$
Thus the wave function of a subsystem has an interesting time dependence. Time appears here in two places: the wave function of the universe depends on $t$ since it evolves according to \Sc's equation. And the configuration of the environment $Y$ also evolves and depends on $t$ as part of the evolving configuration of the universe $Q_t=(X_t,Y_t)$. 

This suggest a rich variety of ways that the wave function of a subsystem might behave in time. Everyone readily believes, and it is in fact the case, that the wave function of a subsystem evolves just as it should for a Bohmian system, according to \Sc's equation for the subsystem, when the subsystem is suitably decoupled from its environment. And it's actually rather easy to see that the wave function of a subsystem collapses according to the usual textbook rules with the usual textbook probabilities in the usual measurement situations. The wave function of the $x$-system thus collapses in just the way wave functions in quantum mechanics are supposed to collapse. This follows more or less directly from standard quantum measurement theory together with the definition of the wave function of the $x$-system and the quantum equilibrium hypothesis \cite{DGZ92}. 

It is a sociological fact, for whatever reason, that even very talented mathematical physicists have a lot of trouble accepting that the wave function of a subsystem  collapses as claimed. We guess that's because people know that collapse in quantum mechanics is supposed to be some really problematical, difficult issue, so they think it can't be easy for Bohmian mechanics either. But it is easy for Bohmian mechanics. And the thing everyone is happy to take for granted, that the wave function of a subsystem  will evolve according to \Sc's equation in the appropriate situations---they do so presumably because nobody says there's a problem getting wave functions to obey \Sc's equation. It's collapse that's the problem. But understanding why the wave function of a subsystem does indeed evolve according to  \Sc's equation when the subsystem is suitably decoupled from its environment is a bit tricky. Nonetheless it is true, though we shall not go into any details here, \z{see} \cite{DGZ92}. 

The main point we wish to have conveyed in this section is that for a Bohmian universe the wave function of a subsystem of that universe, defined in terms of the wave function of the universe and additional resources available to  Bohmian mechanics and absent in \OQT---namely the actual configuration of the environment of the subsystem---behaves exactly the way wave functions in \OQT\ are supposed to behave.

\section{The Wave Function as Nomological}%
The main thing we want to discuss here is the status of the wave function: what kind of thing it is. And what we want to suggest one should think about is the possibility that it's nomological, nomic---that it's really more in the nature of a law than a concrete physical reality. 

Thoughts in this direction might originate when you consider the unusual kind of way in which Bohmian mechanics is formulated, and the unusual kind of behavior that the wave function undergoes in Bohmian mechanics. The wave function of course affects the behavior of the configuration, i.e., of the particles. This is expressed by the guiding equation \eqref{ge}, which in more compact form can be written 
\begin{equation}\label{ge1}
 dQ/dt=v^{\psi}(Q)\,.
\end{equation}
But in Bohmian mechanics there's no back action, no effect in the other direction, of the configuration upon the wave function, which evolves autonomously via \Sc's equation \eqref{se}, in which the actual configuration $Q$ does not appear. Indeed the actual configuration could not appear in \Sc's equation since this equation is an equation also of \OQT\ and in \OQT\ there is no actual position or configuration. That's one point. 

A second point is that for a multi-particle system the wave function $\psi(q)=\psi(\mathbf q_{_1}, \dots,  \mathbf q_{_N})$ is not  a weird field on physical space, its a weird field on configuration space, the set of all hypothetical configurations of the  system. For a system of more than one particle that space is not physical space. What kind of  thing is this field on that space?\footnote{The sort of physical reality to which the wave fucntion corresponds is even more abstract that we've conveyed so far. That's because the wave function, in both \OQT\ and Bohmian  mechanics, is merely a convenient representative of the more physical ``quantum state.'' Two wave functions such that  one is  a (nonzero) scalar multiple of the other represent the same quantum state and are regarded as physically equivalent. Thus the quantum state is not even a field at all, but an equivalence class of fields. It is worth noting that equivalent wave functions define the same velocity \eqref{ge}. They also define, with suitable normalization, the same $|\psi|^2$-probabilities. 

Moreover, for the treatment of identical particles such as electrons in Bohmian mechanics, it is best to regard them  as unlabelled, so that the configuration space of $N$ such particles is not a high dimensional version of a familiar space, like ${\mathbb R}^{3N}$, but is instead the unfamiliar high-dimensional space $\nrd$ of $N$-point subsets of ${\mathbb R}^{3}$. This space has a nontrivial topology, which naturally leads to the possibilities of bosons and fermions---and in two dimensions anyons as well \cite{DGTTZ06}. As a fundamental space it is odd, but not as a configuration space.} 

The fact that Bohmian mechanics requires that one take such an unfamiliar sort of entity seriously  bothers a lot of people. It doesn't in fact bother us all that much, but it does seem like a significant piece of information nonetheless. And what it suggests to us is that you should think of the wave function as describing a law and not as some sort of concrete physical reality. After all \eqref{ge1} is an equation of motion, a law of motion, and the whole point of the wave function here is to provide us with the law, i.e., with the right hand side of this equation. 

Now we've said that rather cavalierly. There are lots of problems with saying it at this point. But before going into the problems let us make a comparison with a familiar situation where nobody seems to  have much of a  problem at all, namely classical Hamiltonian dynamics.

\subsection{Comparison of $\psi$ with the Classical Hamiltonian $H$}
The wave function is strange because it lives on configuration space, for an $N$-particle system a space of dimension 3$N$. Well, there's a space in the classical mechanics of an $N$-particle system that has twice that dimension, its phase space, of dimension 6$N$. On that space there's a function, the Hamiltonian $H = H(q,p)=H(\mathscr{X})$ of the system, and to define the equations of motion of classical mechanics you put $H$ on the right hand side of the equations of motion after suitably taking derivatives. We've never heard anyone complaining about classical mechanics because it invokes a weird field on phase space, and asking what kind of thing is that?  Nobody has any problem with that. Everybody knows that the Hamiltonian is just a convenient device in terms of which the equations of motion can be nicely expressed. 

We're suggesting that you should regard the wave function in exactly the same way. And if you want to have a sharper analogy you can think not of  $\psi$ itself but of something like $\log\psi(q)$ as corresponding to the Hamiltonian $H(\mathscr{X})$. The reason we suggest this is because the velocity in Bohmian mechanics  is proportional to the imaginary part of $\nabla\psi/\psi$ for a scalar wave function, a sort gradient of the log of $\psi$, some sort of derivative, der, of  $\log\psi(q)$, so that \eqref{ge} can be regarded as of the form
$$dQ/dt = \text{der}(\log\psi)\,.$$
Similarly in classical mechanics we have an evolution equation of the form
$$d\mathscr{X}/dt=\text{der\,} H$$
where der $H$ is a suitable derivative of the Hamiltonian. (This is a compact way of writing the familiar Hamiltonian equations $d\mathbf q_k/dt=\partial H/\partial \mathbf p_k$, $d\mathbf p_k/dt=-\partial H/\partial \mathbf q_k$.)

It is also true that both $\log\psi$ and $H$ are normally regarded as defined only up to an additive constant: When you add a constant to $H$ it doesn't change the equations of motion. If you multiply the 
wave function by a scalar---which amounts to adding a constant to its log---the new wave function is generally regarded as physically equivalent to the original one. And in Bohmian mechanics the new wave function defines the same velocity for the configuration, the same equations of motion,  as the original one.

Moreover, with suitably ``normalized'' of choices $\psi(q)$ and $H(\mathscr{X})$, corresponding to appropriate choices of the constants, one associates rather similar probability formulas: In classical statistical mechanics there are the Boltzmann-Gibbs probabilities, given by  $e^{-H/kT}$ when $H$ has been suitably normalized, where $k$ is  Boltzmann's constant and $T$ is the temperature. One thus has that
 $$ \log Prob \propto -H\,.$$
And in quantum mechanics or Bohmian mechanics, with $|\psi|^2$--probabilities, one has that
$$\log Prob \propto \log |\psi|\,.$$
(You probably shouldn't take this last point about analogous probabilities too seriously. It's presumably just an accident that the analogy seems to extend this far.) 

\subsection{$\psi\quad \text{versus} \quad \Psi$}%
There are, however, problems with regarding the wave function as nomological. Laws aren't supposed to be dynamical objects, they aren't supposed to change with time, but the wave function of a system typically changes with time. And laws are not supposed to be things that we can control---we're not God. But the wave function is often an initial condition for a quantum  system. We often, in act, prepare a system in a certain quantum state, that is, with a certain wave function. We can in this sense control the wave function of a system. But we don't control a law of nature. This makes it a bit difficult to regard the wave function as nomological.

But with regard to this difficulty it's important to recognize that there's only one wave function we should be worrying about, the fundamental one, the wave function $\Psi$ of the universe. In Bohmian mechanics, the wave function $\psi$ of a subsystem of the universe is defined in terms of the universal wave function $\Psi$. Thus, to the extent that we can grasp the nature of the universal wave function, we should understand as well, by direct analysis, also the nature of the objects that are defined in terms of it, and in particular we should have no further fundamental question about the nature of the wave function of a subsystem of the universe. So let's focus on the former.

\subsection{The Universal Level}%
When we consider, instead of the wave function of a typical subsystem,  the wave function  $\Psi$ of the universe itself, the situation is rather dramatically transformed. $\Psi$  is not controllable. It is what it is! And it may well not be dynamical either. There may well be no ``$t$'' in $\varPsi$. 

The fundamental equation for the wave function of the universe in canonical quantum cosmology is the Wheeler-DeWitt equation \cite{dW67},
$$\mathscr{H}\varPsi=0\,,$$
for a wave function $\Psi(q)$ of the universe, where  $q$ refers to 3-geometries and to whatever other stuff is involved, all of which correspond to structures on a 3-dimensional space.  In this equation $\mathscr{H}$ is a sort of generalized Laplacian, a cosmological version of a \Sc\ Hamiltonian $H$. And like a typical $H$, it involves nothing like an explicit time-dependence. But unlike \Sc's equation, the  Wheeler-DeWitt equation has on one side, instead of a time derivative of $\Psi$, simply 0. Its natural solutions are thus time-independent, and these are the solutions of the Wheeler-DeWitt equation that are relevant in quantum cosmology.  

That this is so is in fact the {\em problem of time} in quantum cosmology. We live in a world where things change. But if the basic object in the world is a timeless wave function how does change come about? Much has been written about this problem of time.  A great many answers have been proposed. But what we want to emphasize here is that from a Bohmian perspective the timelessness of  $\Psi$ is not a problem. Rather it is just what the doctor ordered. 

The fundamental role of the wave function in Bohmian mechanics is to govern the motion of something else. Change fundamentally occurs in Bohmian mechanics not so much  because the wave function changes but because the thing $Q$  it's governing does, according to a law 
\begin{equation}\label{ge2}
dQ/dt=v^{\varPsi}(Q)
\end{equation}
determined by the wave function. The problem of time vanishes entirely from a Bohmian point of view. And it's just what the doctor ordered because laws are not supposed to change with time, so we don't want the fundamental wave function to change with time. It's good that it doesn't change with time. 

There may be another good thing about the wave function of the universe: it may be unique. It is of course the case that, together with being uncontrollable,  a timeless wave function of our actual universe would be the one wave function that it is. But we mean more than that: While the Wheeler-DeWitt equation presumably has a great many solutions $\Psi$, when supplemented with additional natural conditions, for example  the Hartle-Hawking boundary condition \cite{HH83}, the solution may become unique. And such uniqueness fits nicely with the conception of the wave function as law.

\subsection{Schr\"odinger's Equation as Phenomenological (Emergent)}%
Now we can well imagine someone saying, OK fine, in this Bohmian theory  for the universe stuff changes---particles move, the gravitational field changes, the gravitational metric evolves, whatever. But we know that the most important equation in quantum mechanics, and one of the most important equations in our quantum world, is the time-dependent \Sc\ equation, describing wave functions that themselves change with time. Where does that come from in a theory in which the only fundamental wave function that you have is the timeless wave function  $\Psi$?

But that question has already been answered here, in the last paragraph of Section \ref{cwf}. If you have a wave function of the universe obeying \Sc's equation, then in  suitable situations, those in which a subsystem is suitably decoupled from its environment (and the Hamiltonian $\mathscr{H}$ is
of the appropriate form),  the wave function $\psi_t(x)=\Psi\!(x,Y_t)$ of the subsystem
will evolve according to Schr\"odinger's equation for that subsystem.  For $\Psi$ not depending upon time, the wave function of the subsystem inherits its time-dependence from that of the configuration $Y_t$ of the environment. And the crucial point here is that a  solution $\Psi_t$ of \Sc's equation can be time-independent. These are the solutions $\Psi_t$ that are the same wave function $\Psi$ for all t, the solutions for which $\partial \Psi_t/\partial t$ is 0 for all $t$, corresponding precisely to solutions of the  Wheeler-DeWitt equation.

But in this situation the time-dependent \Sc\ evolution \eqref{se} is not fundamental. Rather it is emergent and phenomenological, arising---as part of a good approximation for the behavior of suitable subsystems---from a Bohmian dynamics \eqref{ge2} for the universe given in terms of a suitable wave function of the universe, one which obeys the Wheeler-DeWitt equation. 

And even this time-independent equation might not be fundamental---that it appears to be might be an illusion. What we have in mind it this: We've got a law of motion involving a vector field  $v^{\Psi}$ (the right-hand-side of a first-order equation of motion), a vector field that can be expressed in terms of $\Psi$. If  $\Psi$ is a nice sort of wave function it might obey all sorts of nice equations, for example the Wheeler-DeWitt equation or something similar. From a fundamental point of view, it might be a complete accident that $\Psi$ obeys such an equation. It might just happen to do so. The fact that the equation is satisfied might have nothing to do with why the fundamental  dynamics is of the form \eqref{ge2}. But as long as $\Psi$ does satisfy the equation, by accident or not, all the consequences of satisfying it follow. 

So it could turn out, at the end of the day, that what we take to be the fundamental equation of quantum theory, \Sc's equation, is not at all fundamental for quantum theory, but rather is an emergent and  accidental equation. 
\subsubsection{Two Transitions}%
We'd like to focus a bit on the change of perspective that occurs when we make the transition from \OQT, which seems to involve only the wave function $\psi$, to (conventional) Bohmian mechanics, which is usually regarded as involving two types of physical entities, wave functions $\psi$ and the positions of particles, forming a configuration $Q$, to universal Bohmian mechanics, where the wave function $\Psi$ is taken out of the category of concrete physical reality and into that of law, so you've got just $Q$ as describing elements of physical reality: 

\parbox{1in}{\bc OQT\\$\psi$\ec}$\longrightarrow$\parbox{1in}{\bc BM\\$(\psi,Q)$\ec}$\longrightarrow$\parbox{1in}{\bc UBM\\$Q$\ec}\\
You start with just $\psi$, you end with just $Q$. 

And our original question, about the wave function---What kind of thing is that?---is rather dramatically transformed when we make this transition to the universal level, since we're then asking about a very different object, not about a wave function of a subsystem of the universe but about the universal wave function, 

\parbox{2in}{\bc?\\
?$\quad\psi\quad$?\\
?\ec}
$\ \longrightarrow${
\ \ \parbox{2in}{\bc?\\
?$\quad$ $\varPsi$\normalsize$\quad$?\\
?\ec}}\\
which is actually, so we are supposing, just a way of representing the law of motion. So, now we may ask,  does any kind of question about $\Psi$ remain? 

Here's one question: Why should the motion be of the form \eqref{ge2},  involving $\Psi$ in the way that it does? Why should the law of motion governing the behavior of the constituents of the universe be of such a form that there is a wave function in terms of which the motion can be compactly expressed? We think that's a good question. And of course we have no definitive answer to it. But an answer to this question would provide us with a deep understanding of why our world is quantum mechanical. 

The view that the wave function is nomological has another implication worth considering. This is connected with the question of how we ever come to know what the wave function of a system is. There must be some algorithm that we use. We don't directly see wave functions. What we see (more directly) are particles, at least from a Bohmian perspective. We should read off from the state of the primitive ontology, whatever it may be, what the relevant wave function is. There should be some algorithm connecting the state of the primitive ontology, for Bohmian mechanics the relevant configuration $Q(t)$ over, say, some suitable time interval, with the relevant wave function. 

Now let's go to the universal level. You might think there should be some algorithm that we can use to read off what the universal wave function is from the state of the primitive ontology of the universe, whatever that may be. But in fact we kind of doubt there is any such algorithm. So far as we know, nobody has proposed any such algorithm. And from the point of view of the wave function $\Psi$ being nomological, you wouldn't expect there to be any such algorithm. That's because if the wave is nomological, specifying the wave function amounts to specifying the theory. You wouldn't expect there to be an algorithm for theory formation. 
\subsection{Nomological versus Nonnomological}
Now, we can imagine---and in fact we're quite sure---that many physicists would respond to the question about whether or not the wave function is fundamentally nomological with a big ``Who cares? What difference does it make?'' 

Well, we think it does matter. Being nomological has important implications. Laws should be simple. If we believe that the wave function of the universe is nomological, this belief should affect our expectations for the development of physics. We should expect somehow to arrive at physics in which the universal wave function involved in that physics is in some sense simple---while presumably having a variety of other nice features as well. 

Now simplicity itself is sort of complicated. There are a number or varieties of simplicity. For example, the universal wave function could be simple in the sense that it has a simple functional form---that it's a simple function of its arguments. That's one possibility. Another, quite different, is that it could be a more or less unique solution to a simple equation.  Or, a similar kind of thing, it could more or less uniquely satisfy some compelling principle, maybe a symmetry principle. 

For example, Stefan Teufel and one of us explored the possibility that a symmetry principle expressing a sort of quasi-4-diffeomorphism invariance would imply an evolution of 3-geometries governed by a universal wave function \cite{GT01}. In technical terms, we demanded that the vector field on super-space defining the relevant motion form a representation of the ``Dirac algebra''  \cite{Dir64}, a sort of algebra, sort of corresponding to 4-diffeomorphism invariance. That puts very strong constraints on the theory. In fact the constraints are so strong that it seems that the only possibilities correspond to classical general relativity, with nothing genuinely quantum mechanical arising. 

That's for pure gravity. It's not clear what would happen if matter degrees of freedom were included in the analysis. So one could always hope that if matter were included in the story and you played a similar game you would thereby end up with quantum mechanics as the only (reasonable) possibility. But that is highly speculative---only a hope and a prayer. 

\subsection{Relativistic Bohmian Theory}
We want to say a bit about Lorentz invariance and a problem that arises in connection with it. It's widely said---and it's natural to think---that you can't have a Lorentz invariant Bohmian theory. That's basically because of the crucial role played in such a theory by the configuration of the system: the positions of its particles---or the detailed description of the primitive ontology of the theory, whatever that may be---{\em  at a given time.} 

Now you can also consider configurations determined, not by a $t=$ constant hypersurface but by a general space-like hypersurface. For example, for a particle ontology, the configuration corresponding to such a surface would be given by the space-time points  on the surface at which the world-lines of the particles cross the surface, for $N$ particles, $N$ points. So if in fact you had somehow at your disposal a Lorentz invariant foliation of space-time into space-like hypersurfaces, you could play a Bohmian game and define a Bohm-type dynamics for the evolution of configurations defined in terms of that foliation. In this way one could obtain a Lorentz invariant Bohmian theory \cite{DGMZ99}.

To actually have such a thing the best possibility is perhaps the following: You have a Lorentz invariant rule for defining in terms of the universal wave function a foliation of space-time, a covariant map fol from wave functions to foliations:
\bc{
(Lorentz) covariant map $\Psi \stackrel{\text{\normalsize fol}}{\longrightarrow}  \mathscr F=\mathscr F(\Psi)$\,.}\ec
(For the map to be {\em covariant} means that the diagram
$$
\begin{array}{rcl}
\varPsi\, & \stackrel{\text{\normalsize fol}}{\longrightarrow} & \mathscr F\\
\Lambda_{\text{g}} \Big\downarrow \:\: && \Big\downarrow \text{ g}\\
\varPsi_{\text{g}} & \stackrel{\text{\normalsize fol}}{\longrightarrow} & {\mathscr F}_{\text{g}}
\end{array}
$$
is commutative. Here g is any Lorentz transformation, on the right acting naturally on the foliation by moving the points on any leaf of the foliation, and hence the leaves themselves and the foliation itself,  around according to g, while $\Lambda_{\text{g}}$ is the action of g on wave functions, given by a representation $\Lambda$ of the Lorentz group.)

Lorentz invariant Bohmian theories formed in this way, by utilizing such  a covariant foliation map, have the virtue of being seriously Lorentz invariant. The point here is that any theory can be made Lorentz invariant in a trivial nonserious way by introducing suitable additional space-time structure beyond the Lorentz metric. The question then arises as to what kinds of structure are unproblematic. James Anderson has addressed this question by distinguishing between absolute and dynamical structures, and identifying the serious Lorentz invariance of a theory with the nonexistence in the theory of any additional absolute structures \cite{And67}. 

What exactly these are is not terribly relevant for our purposes here. That's because, for the sort of theory proposed here, what seems to be additional space-time structure, namely the foliation, is not an additional structure at all, beyond the wave function.  To the extent that the wave function is a legitimate structure for a Lorentz invariant theory---and this is generally assumed to be the case---so are covariant objects defined solely in terms of the wave function.

Here are some examples of possibilities for covariant foliations. You could form a typical quantum expectation in the Heisenberg picture, involving the universal wave function and some sort of operator-valued Fermi field $\psi(x)$. The simplest such object is perhaps
$$
j_{\mu}(x)=\left\langle\Psi\left|\bar{\psi}(x)\gamma_{\mu}\psi(x)\right|\Psi\right\rangle\,,
$$
involving the Dirac matrices $\gamma_{\mu}$, defining a time-like vector field on space-time. You could also put suitable products in the middle to form tensors of various ranks. Ward Struyve  \cite{prep11} has suggested using the stress energy tensor
$$
t_{\mu\nu}(x)=\left\langle\varPsi\!\left|T_{\mu\nu}(x)\right|\!\varPsi\right\rangle
$$
and integrating that over space-like hypersurfaces to obtain a time-like vector (that in fact does not depend upon the choice of surface). There are a variety of such proposals for extracting from the wave function a vector field on space-time in a covariant manner. 
And a vector field on space-time is just the sort of thing that could define a foliation, namely into hypersurfaces orthogonal to that vector field, so that we have the following scheme for a map fol:
$$
\Psi\to j_{\mu}\leadsto\mathscr F
$$

Now Struyve's proposal works as is, but for other proposals you would have to do a lot of massaging to get  the scheme to work: In order to define a foliation the vector field would have to be what is called ``in involution.'' That can be achieved, but in so doing you would like the resulting vector field to remain time-like (so that the corresponding foliation would be into space-like hypersurfaces), and that is certainly not automatic. 

The bottom line is that there is lots of structure in the universal wave function, enough structure certainly to typically  permit the specification of a covariant rule for a foliation. 

\subsection{Wave Function as Nomological and Symmetry}
But there is a problem: there is a conflict between the wave function being nomological and symmetry demands. The problem arises from the difference between having an action of the Lorentz group $G$ (or whatever other symmetry group we have in mind)  on the Hilbert space ${\cal H}$ of wave functions (or on a suitable subset of ${\cal H}$, the domain of the foliation map)---which is more or less all that is usually required for a Lorentz invariant theory---and having the trivial action, always carrying $\Psi$ to itself: the difference between an action of $G$ on ${\cal H}$ and the $G$-invariance of $\varPsi$. If the universal wave function
represents the law, then that wave function itself, like an invariant law, should be $G$-invariant. (Actually, any change of the wave function that leaves the associated velocity vector field alone would be fine, for example multiplication by a constant scalar, but we shall for simplicity ignore this possibility here.)

It's not hard to see that that's incompatible with the covariance of the foliation map. No foliation can be Lorentz invariant, since there is always some Lorentz transformation that will tilt some of its leaves, at least somewhere. But if $\Psi$ is g-invariant so must be any foliation associated with $\Psi$ in a covariant manner. Thus a Lorentz invariant wave function $\Psi$ can't be covariantly associated with a foliation. 

This is in sharp contrast with the situation for a generic wave function of the universe. Such a wave function will not be symmetric, and there is no obstacle to its being in the domain of a covariant foliation map. But if the universal wave function is nomological, it is not  generic, and it must be too symmetric to permit the existence of a covariant foliation map.

\subsubsection{Possible Resolutions}

There seems to be a conflict between (i) having a Bohmian quantum theory, (ii) the universal wave function for that theory being nomological, and (iii) fundamental Lorentz invariance. Something, it would seem, has to give. And from a Bohmian point view the thing that gives wouldn't be the Bohmian part.   

Here are some possible resolutions that would allow us to continue to regard the wave function as nomological. You could abandon fundamental Lorentz invariance, as many people have suggested. Another possibility is to make use of a Lorentz invariant foliation, but not one  determined by the wave function but rather  defined in terms of additional dynamical structure beyond the wave function, for example some suitable time-like vector field definable from the primitive ontology or perhaps transcending the primitive ontology, or something like a ``time function'' in general relativity, defined in terms of the gravitational metric and stuff in space-time. Or, the most likely possibility: something nobody has thought of yet.

And, of course, there is the possibility that  we will  have to abandon our attempt to regard the wave function as nomological. Many, for example Travis Norsen,  would then insist that the wave function be eliminated in favor of something like exclusively local beables \cite{Nor09}. That's not how we feel. If it should turn out that the wave function can't be regarded as nomological---because it's too complicated or whatever---our reaction would probaby be: OK, that's just the way it is. It's not nomological but something different. 

In fact, we think in fact that if someone gave us a Bohmian kind of theory, involving a complicated collection of  exclusively local beables, and then someone else pointed out to us that the complicated  local beables can be repackaged into a simple mathematical object of a nonlocal character---like a wave function on configuration space---our reaction would likely be that we would prefer to regard the wave function in that simpler though more unfamiliar way, just because of its mathematical simplicity. 

\subsection{$\psi$ as Quasi-Nomological}
Suppose we accept that the universal wave function $\Psi$ is nomological. What then about the status of the wave function $\psi$ of subsystems of the universe---the wave functions with which we're normally concerned in applications of quantum theory? To this question we have several responses.

Our first response is this: You can decide for yourself. We are assuming the status of $\Psi$ is clear.  The status of the primitive ontology is certainly clearer still. Therefore, since $\psi$ is defined in terms of $\Psi$ and the primitive ontology (specifically, the configuration $Y$ of the environment), the status of $\psi$ must follow from an analysis of its definition.  

We don't insist that everyone would agree on the conclusion of such an analysis. It may well be that different philosophical prejudices will tend to lead to different conclusions here. Our point is rather that once the status of the wave function of the universe has been settled, the question about the status of $\psi$ is rather secondary---something about which  one might well feel no need to worry.

Be that as it may, we would like to regard $\psi$ as quasi-nomological. We mean by this that while there are serious obstacles to regarding the wave function of a subsystem as fully nomological,  $\psi$ does have a nomological aspect in that it seems more like an entity that is relevant to the behavior of concrete physical reality (the primitive ontology) and not so much  like a concrete physical reality itself.

But we can say more. The law governing the behavior of the primitive ontology of the universe naturally implies a relationship between the behavior of a subsystem and the configuration of its environment. It follows from its definition \eqref{cwf1} that the wave function of the subsystem captures that aspect of the environment that expresses this relationship---that component of the universal law that is relevant to the situation at hand, corresponding to the configuration of the environment.

\section{The Status of the Wave Function in Quantum Theory}
Let's return to the possibilities for the wave function, mentioned in Section 1. It could be nothing. While not exactly nothing, it could be merely subjective or epistemic, representing our information about a system. Or it could be something objective. If it is objective, it could be material or quasi-material, or it could be nomological, or at least quasi-nomological. 

Rather than deciding in absolute terms which of these possibilities is correct or most plausible---concerning which our opinion should be quite clear---we conclude by stressing that one's answer to this question should depend upon one's preferred version of quantum theory. Here are some examples: 
\begin{itemize}
 \item In \OQT\ the wave function is  quasi-nomological. It governs the results of quantum ``measurements''---it provides statistical relationships between certain macroscopic variables.
\item In Everett the wave function is quasi-material. After all, there it's all that there is. In Everett there's only the wave function. (We say ``quasi-material'' here instead of plain material since in Everett the connection between the wave function and our familiar material reality is not at all straightforward. In fact for some Everettians part of the appeal of their approach is the extensive conceptual functional analysis that it requires\z{, see, e.g.,} \cite{wallace03}.) 
\item In Bohmian mechanics as we understand it, as well as in decoherent or consistent histories and in casual set theory, the wave function is either nomological or quasi-nomological: In these theories the wave function governs the behavior of something else, something more concretely physical.
\item However, in David Albert's version of Bohmian mechanics \cite{albert1}, in which what we call configuration space is in fact a very high-dimensional {\em physical} space, on which the wave function lives as a physical field, the wave function is material or quasi-material.
\item In GRW theory  \cite{GRW86} or CSL \cite{GPR90}---the theory of continuous spontaneous locali\-zation---the wave function is quasi-material, since there it either is everything or at least determines everything.
\item And in  the quantum information approach to quantum theory, the wave function is quasi-subjective---``quasi'' because quantum information theorists differ as to how subjective it is. 
\end{itemize}
This means that if you want to grasp the status of the wave function in quantum theory, you need to know exactly what it is that quantum theory says. If you're not clear about quantum theory, you shouldn't be worrying about its wave function.

We have suggested  seriously considering the possibility that the wave function is nomological. One psychological obstacle to doing so is this:  It seems to be  an important feature of wave functions that they are variable and that this variability---from system to system and not just over time---leads to the varieties of different behaviors that are to be explained by quantum theory.  But the behavior of the primitive ontology of a Bohmian theory, and all of the empirical consequences of the theory, depend on the universal wave function only via the one such wave function that exists in our world and not on the various other universal wave functions that there might have been. The variability we see in wave functions is that of wave functions of subsystems of the universe. This variability originates in that of the environment $Y$ of the subsystem as well as that of the choice of subsystem itself. So this variability does  not conflict with regarding the wave function as fundamentally nomological, but rather is explained by it.

\end{document}